\documentclass[lettersize,journal]{IEEEtran}
\usepackage[commandnameprefix=always]{changes}

\usepackage{amsmath,amsfonts}
\usepackage{algorithmic}
\usepackage{algorithm}
\usepackage{array}
\usepackage{graphicx}
\usepackage[caption=false,font=normalsize,labelfont=sf,textfont=sf]{subfig}
\usepackage{cite}
\usepackage{textcomp}
\usepackage{url}
\usepackage{verbatim}
\usepackage{stfloats}
\usepackage{comment}
\usepackage{svg}
\usepackage{fancyhdr} 

\fancypagestyle{plain}{
}
\begin{document}

\title{CCAT: Optical Responsivity, Noise, and Readout Optimization of KIDs for Prime-Cam}

\author{Eliza Gazda, Quintin Meyers, Steve K. Choi,  James R. Burgoyne, Scott Chapman, Cody J. Duell, Anthony I. Huber, Inchara Jagadeesh, David Faulkner Katz, Ben Keller, Lawrence T. Lin, Paul Malachuk, Michael D. Niemack, Darshan A. Patel,  Gordon J. Stacey,  Benjamin J. Vaughan, Eve M. Vavagiakis, Samantha Walker, Yuhan Wang, Ruixuan(Matt) Xie

\thanks{E. Gazda, S. K. Choi, I. Jagadeesh, D. Faulkner Katz, P. Malachuk, Q. Meyers are with the Department of Physics and Astronomy, University of California, Riverside, CA 92521, USA. (e-mail: egazda@ucr.edu) 

C. J. Duell, B. Keller, L. T. Lin, M. D. Niemack, D. A. Patel,  G. J. Stacey, B. J. Vaughan, S. Walker, and Y. Wang are with the Department of Physics, Cornell University, Ithaca, NY 14853, USA. 

J. R. Burgoyne is with the Department of Physics and Astronomy, University of British Columbia, Vancouver, BC V6T 1Z1, Canada

S. Chapman is with the Department of Physics and Atmospheric Science, Dalhousie University, Halifax, NS B3H 4R2, Canada.

A. I. Huber, R. Xie are with Department of Physics and Astronomy, University of British Columbia, Vancouver, BC, Canada.

E. M. Vavagiakis is with the Department of Physics, Duke University, Durham, NC, 27704, USA, and the Department of Physics, Cornell University, Ithaca, NY 14853, USA.
}}



\maketitle
\thispagestyle{plain}

\begin{abstract}
The Prime-Cam instrument on the Fred Young Submillimeter Telescope (FYST) at the CCAT Observatory will conduct sensitive millimeter to submillimeter surveys for a range of astrophysical and cosmological sciences. Prime-Cam will use kinetic inductance detectors (KIDs) sensitive to multiple frequency bands spanning 280–850 GHz. With over 100,000 sensors under development, these KID arrays will soon form the largest submillimeter focal plane ever built. With fixed microwave tones probing amplitude and phase modulations in the KIDs due to incoming radiation, challenges arise in determining the optimal readout settings, especially under varying atmospheric loading. Realizing the science goals of FYST requires operating the detectors at optimal performance and determining accurate responsivities, which depend on readout tone placement and power. To address these challenges, we present laboratory measurements of sample pixels from the 280 GHz TiN and Al arrays using a blackbody cold load to simulate observing conditions. These measurements probe detector responsivity and noise across varying optical loading, tone power, and tone placement, providing the foundation to guide in situ calibration and operation of the $>$100,000 KIDs. We characterize detector sensitivity via the Noise Equivalent Power (NEP) as a function of readout tone power and placement, and measure the impact of detuning due to varying optical power on the NEP. Our test setup and methodology will inform the commissioning of Prime-Cam, in situ detector calibration procedures, the cadence of probe tone resetting, and potential design refinements for future arrays, supporting FYST’s planned first light in 2026.

\end{abstract}


\section{Introduction}
The CCAT Observatory's Fred Young Submillimeter Telescope (FYST) is a next generation submillimeter telescope located at a high-altitude site (5,600\,m) on Cerro Chajnantor in the Atacama Desert. Designed to tackle key questions in cosmology and astrophysics, FYST will enable studies of cosmic inflation, galaxy evolution, and transient phenomena \cite{CCAT2022_Science}.  FYST is scheduled for first light in 2026. With a wide field of view and optimized optical design, FYST is well suited for conducting large area surveys at submillimeter wavelengths \cite{Parshley2018,Parshley2022}.

The first-generation receiver on FYST is Prime-Cam \cite{Vavagiakis2018,Choi2020}, a modular instrument designed for simultaneous observations across multiple frequency bands. The complete Prime-Cam focal plane will consist of seven modules, in total populated with over 100,000 Kinetic Inductance Detectors (KIDs). The nominal design includes broadband, polarization-sensitive modules targeting 280, 350, and 850\,GHz, along with imaging spectrometer modules in the 210--420\,GHz band. With photon-noise-limited performance demonstrated \cite{duell2024_ccat_tin_al_comparison, vaskuri2025_ccat_al_mkids, Hubmayr2015}, KIDs were chosen for Prime-Cam due to their natural frequency multiplexing, which enables large scale focal planes with simplified and efficient readout systems\cite{Duell2020}. However, the sensitive dependence of detector performance and responsivity on optical loading, tone power, and tone placement pose a challenge for achieving optimal performance and accurate calibration across many KIDs, which we address through the laboratory characterization measurements presented here.

In this work, we focus on the laboratory characterization of 280\,GHz sample detectors fabricated from two different materials: titanium-nitride (TiN) and aluminum (Al). These detectors were fabricated on wafer edges, alongside the full 280\,GHz arrays, which will be among the first arrays deployed on Prime-Cam. Furthermore, the two materials allow for a comparison of any variations in performance. TiN and Al exhibit different detector characteristics, including nonlinear responsivity for Al and distinct noise behavior, with TiN typically having higher levels of $1/f$ noise. Using a blackbody cold load, we measure the optical responsivity across a range of input powers to evaluate performance and linearity under controlled conditions. From these measurements, we determine the Noise Equivalent Power (NEP) under different optical loadings and readout configurations. Our results show that material choice, optical loading, tone placement and readout/tone power setting all contribute to responsivity and NEP behavior, underscoring the need for detailed characterization and periodic recalibration to ensure detectors meet the sensitivity and stability goals for observations on FYST.

\begin{figure*}[htbp]
\centering
 \includegraphics[width=1\textwidth]{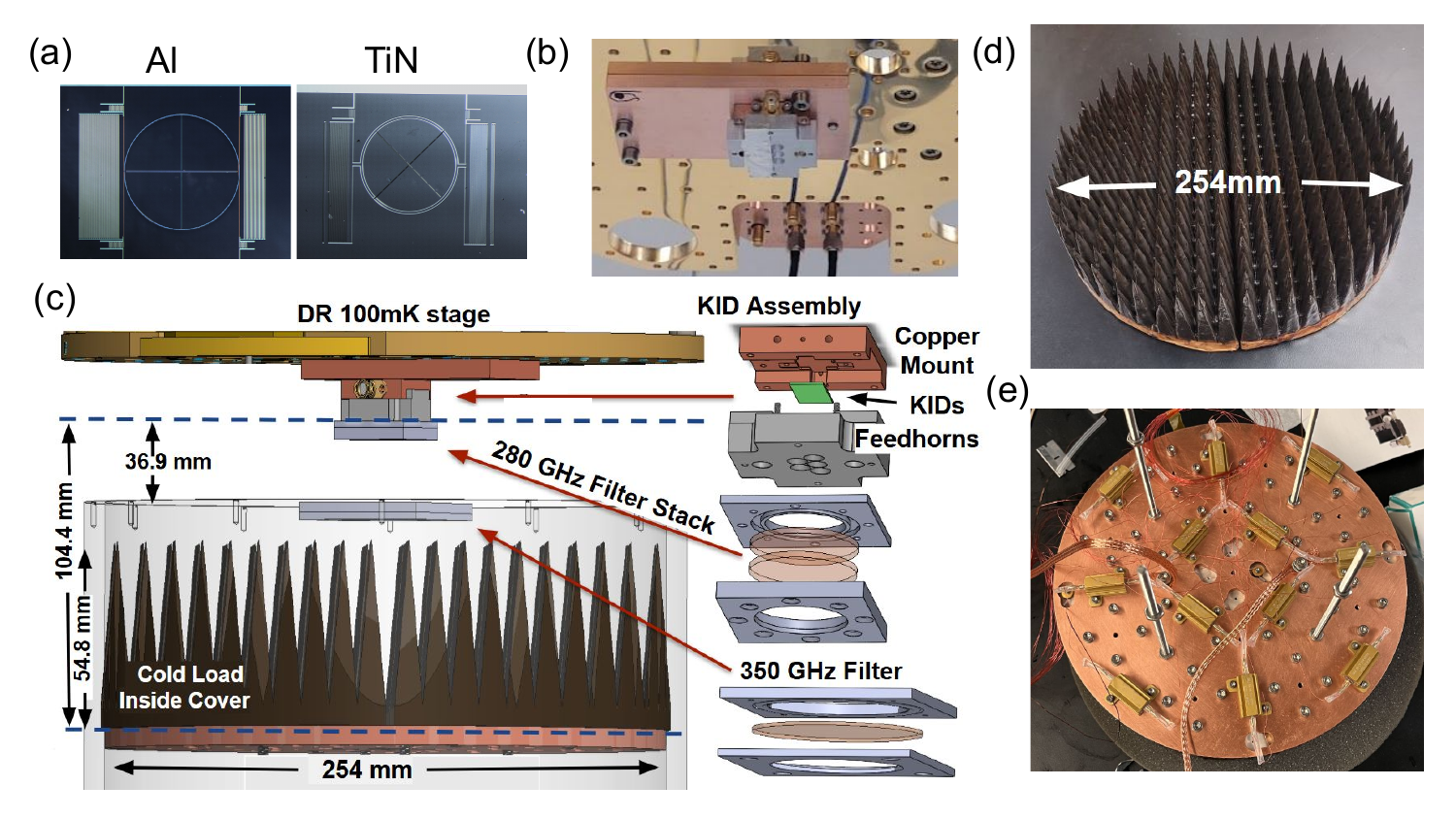}

\caption{Experimental setup. (a) Al and TiN (280\,GHz) sample pixels \cite{Austermann2018, vaskuri2025_ccat_al_mkids}. The central cross and plus- shaped features are inductors, which double as absorbers, while the rectangular structures are interdigitated capacitors. Together, these elements form the LC resonant circuit of each detector. (b) TiN sample array mounted on the 100 mK stage inside the dilution refrigerator (DR). (c) Complete detector, filter stack, and cold load setup inside the DR. (d) Cold load assembly mounted to 4\,K stage. (e) Bottom of the cold load including temperature sensors and heaters controlled by a Lakeshore 325.}
\vspace{-.5em}
\label{fig:setup}
\end{figure*}

\section{Instrument and Test Setup}

The set of sample KIDs tested in this work include 280\,GHz sample detectors fabricated from TiN thin films and Al thin films developed by NIST for submillimeter applications \cite{vaskuri2025_ccat_al_mkids,duell2024_ccat_tin_al_comparison}. Sample detectors are shown in Fig. \ref{fig:setup}a. A KID is composed of a superconducting LC resonator\cite{DayKidz}. Incoming photons of sufficiently high energy change the kinetic inductance of a superconducting film, shifting each detector's resonant frequency and modulating the phase of the probe tones, which allow for sensitive, frequency domain multiplexed measurements of millimeter- and submillimeter-wavelength radiation. The transmission lines of the sample KID arrays are wirebonded to custom-designed printed circuit boards (PCBs). The 280\,GHz pixels are aligned to waveguides with the copper box and the detector chip is aligned via bumpers in the copper box that align the edges of the chip. Optical coupling is provided through conical feedhorns due to their ease of machining. This differs from the feedhorns used in the full arrays, which use a more optimized spline-section design as described in Refs.~\cite{vaskuri2025_ccat_al_mkids,duell2024_ccat_tin_al_comparison}. All components are mounted in a custom aluminum enclosure (Fig.~\ref{fig:setup}c), which provides optical coupling and radiation shielding. 

To define the optical passband and suppress out-of-band radiation, two low-pass metal-mesh filters with measured cutoff frequencies of 288\,GHz and 330\,GHz are mounted at the entrance of the enclosure. The cold load is enclosed in an aluminum shell with a defined aperture, where an additional low-pass filter with a cutoff at 370\,GHz is mounted. These filters combined with the waveguide create an effective passband from 237 GHz to 288 GHz, with the peak transmission level around 0.9. To minimize potential light leaks, the filter stacks are taped around the edges to eliminate gaps between the filters and their mounts. As shown in Fig.~\ref{fig:setup}c, all filters are mounted in custom aluminum holders designed to provide heat sinking to the corresponding temperature stages. Sensitivity to radiation below the band of interest is further suppressed by the detector waveguide geometry, reducing the likelihood of significant red-leak contributions.


Each TiN detector has a superconducting critical temperature (Tc) of approximately 1.1\,K, while the Al detectors have a Tc of 1.4\,K. These values were measured during fabrication at NIST using resistance versus temperature measurements. The entire assembly is mounted inside a dilution refrigerator (Bluefors LD-400), as shown in Fig. \ref{fig:setup}b, which reaches base temperatures below 100\,mK. The test setup also includes an external magnetic shield at 300\,K outside of the cryostat. 

The KID readout is performed using a custom RF system based on a Xilinx RFSoC platform (ZCU111) \cite{Sinclair2022}. The RF chain includes cryogenic attenuators, low noise amplifiers (LNAs), and highpass filters, all optimized for the KID readout frequency band from 500\,MHz to 1\,GHz. This configuration allows for frequency domain multiplexed readout of multiple resonators simultaneously with high dynamic range and low added noise.

To provide controlled and uniform optical loading, we use a blackbody cold load, shown in Fig.~\ref{fig:setup}d, machined from aluminum into a pyramidal geometry and coated with a thin layer of Eccosorb CR-110, a microwave absorber with high emissivity in the submillimeter band. Each pyramid has a height of 54.5\,mm and a base width of 12.7\,mm, and the overall structure has a diameter of 254\,mm, which fills the detector beam. The cold load is attached to the 4\,K bottom shell plate with steel support rods to provide the necessary mechanical stability, while its base is thermally heat sunk using copper braids with a controlled thermal conduction to the bath. The geometry and high emissivity of the Eccosorb coating ensure blackbody emission within the detector passbands \cite{Hemmati:85}. The cold load is temperature controlled with resistive heaters which are thermally coupled to the back of the copper plate, as shown in Fig. \ref{fig:setup}e and actively controlled via a PID feedback loop. The thermal time constant of the system is such that the cold load typically requires $\sim$20–30 minutes to reach equilibrium following a change in the temperature setpoint. Once thermal equilibrium is achieved, the temperature is stable at the setpoint to within $\pm 0.05$\,K over the duration of an optical measurement, allowing repeatable and well-controlled modulation of the optical loading for detector characterization.

\section{Method, Data, and Results}

We characterize detector responsivity and noise over a range of three operating conditions: the readout power (RF attenuation), the frequency placement of the readout tone relative to each detector’s resonance, and the optical power incident on the detectors.
\begin{figure}[!tbp]
    \centering
    \includegraphics[width=0.22\textwidth]{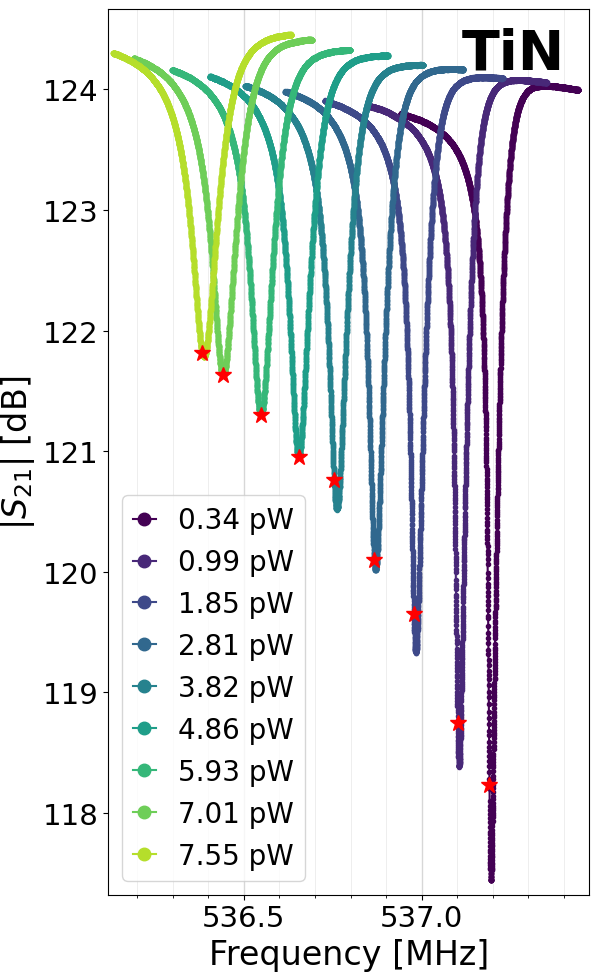}
    \includegraphics[width=0.22\textwidth]{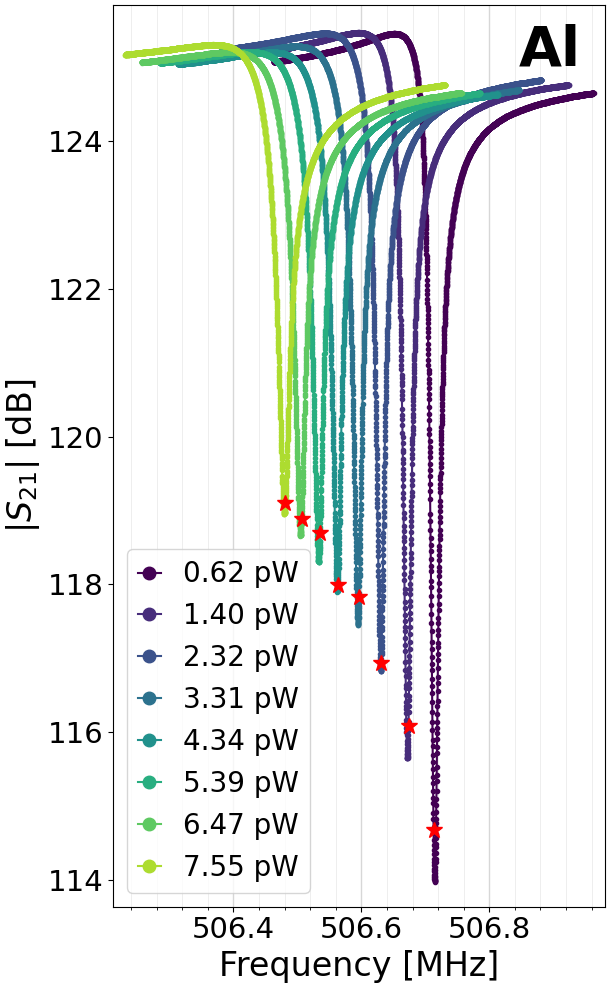}
    \caption{Example $|S_{21}|$ magnitude responses vs frequency of two resonators under multiple optical loading levels. The left plot shows a resonator from the TiN sample array and the right plot shows a resonator from an Al array, both at a readout power of $3$~dB below the bifurcation point when cold load is set to 4\,K. Red stars mark the tone placements where $d\phi/df$ reaches maximum, used during data acquisition. Increasing optical power shifts the resonator frequency to lower values and decreases the quality factor.
    \vspace{-1em}
}
    \label{fig:s21_optical_power}
\end{figure}

In the first set of measurements, we adjust the cold load temperature to vary the optical loading on the detectors and measure forward transmission (S21) sweeps for KID resonant frequencies, phase response, and timestreams at each setting. Next, we change the cold load temperature in small increments while fixing the readout tone frequency. This second measurement is used to examine detector responsivity and noise as a function of tone position offset to assess the potential impact of atmospheric loading drifts on detector calibration and performance under fixed tone readout. Finally, we vary the readout power to determine bifurcation points at high powers and to quantify noise levels at smaller, suboptimal powers.
\begin{figure}[!tbp]
    \centering

    \includegraphics[width=0.43\textwidth]{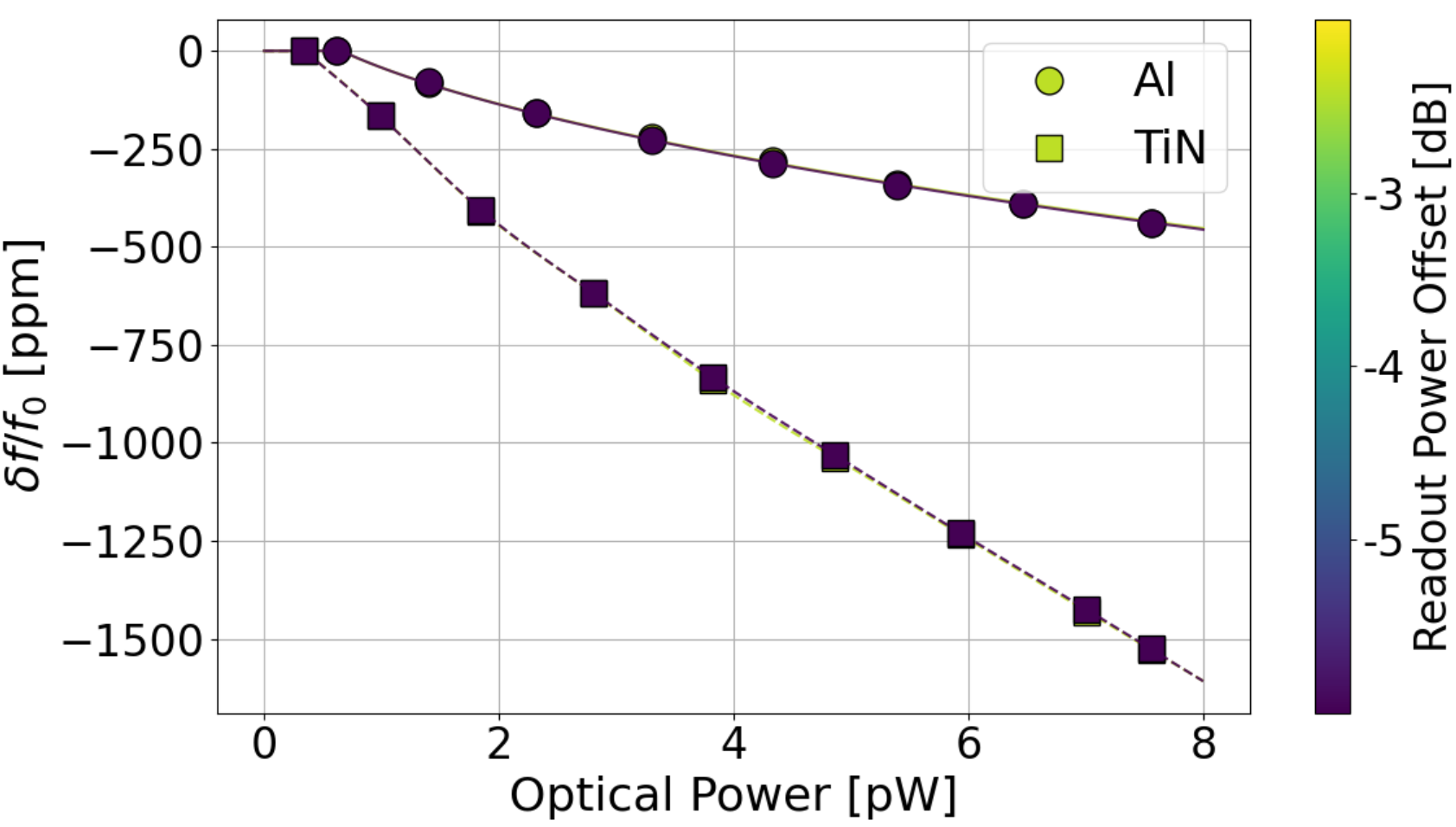}

    \caption{Fractional frequency shift ($\delta f/f_0$) as a function of optical loading for a representative set of 280\,GHz KIDs, used to extract optical responsivity. Readout powers are relative to bifurcation level (at 0\,dB) for resonators.}
    \vspace{-1em}
    \label{fig:responsivity_vs_optical_power}
\end{figure}
\begin{figure}[!tbp]
    \centering

    \includegraphics[width=0.22\textwidth]{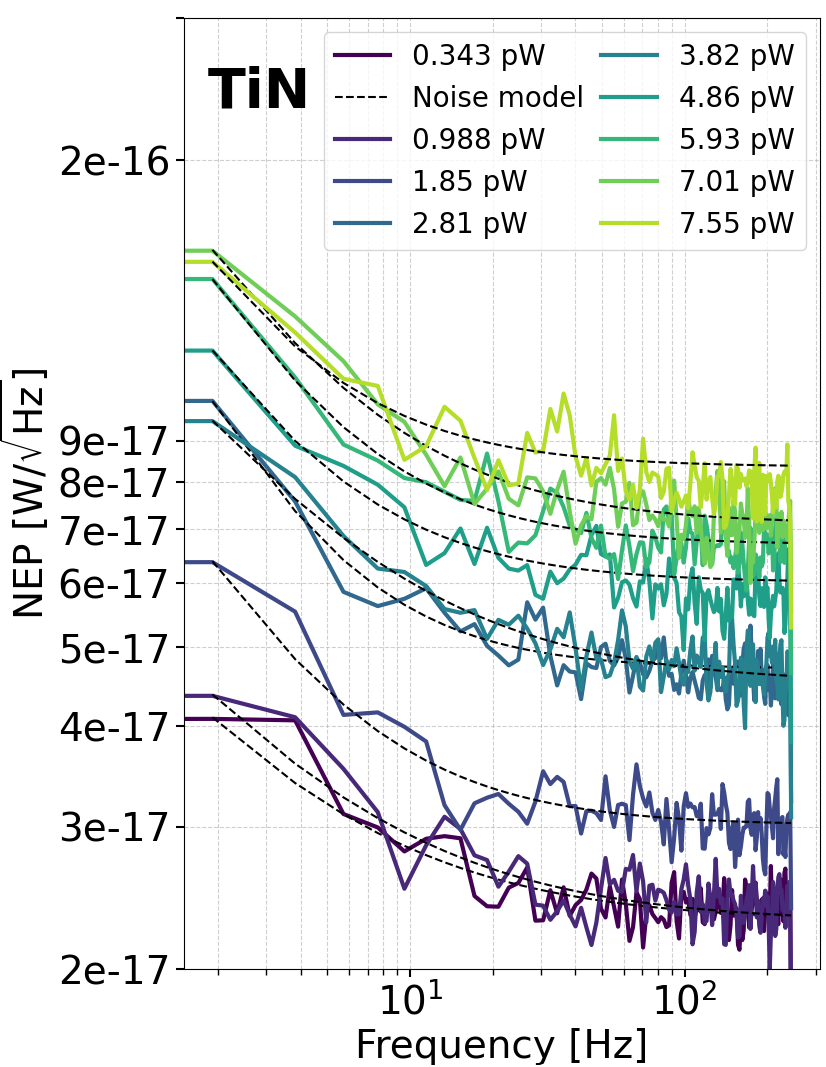}
    \includegraphics[width=0.22\textwidth]{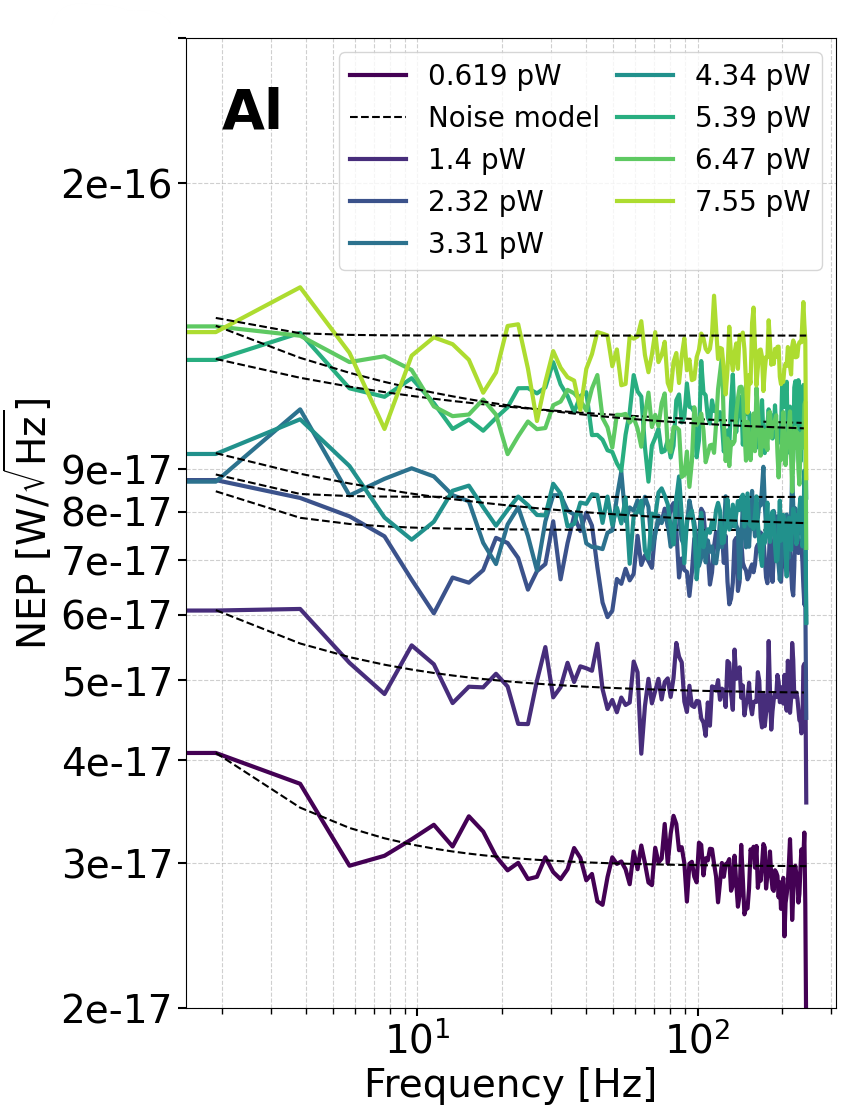}

    \vspace{0.2cm}

    \includegraphics[width=0.4\textwidth]{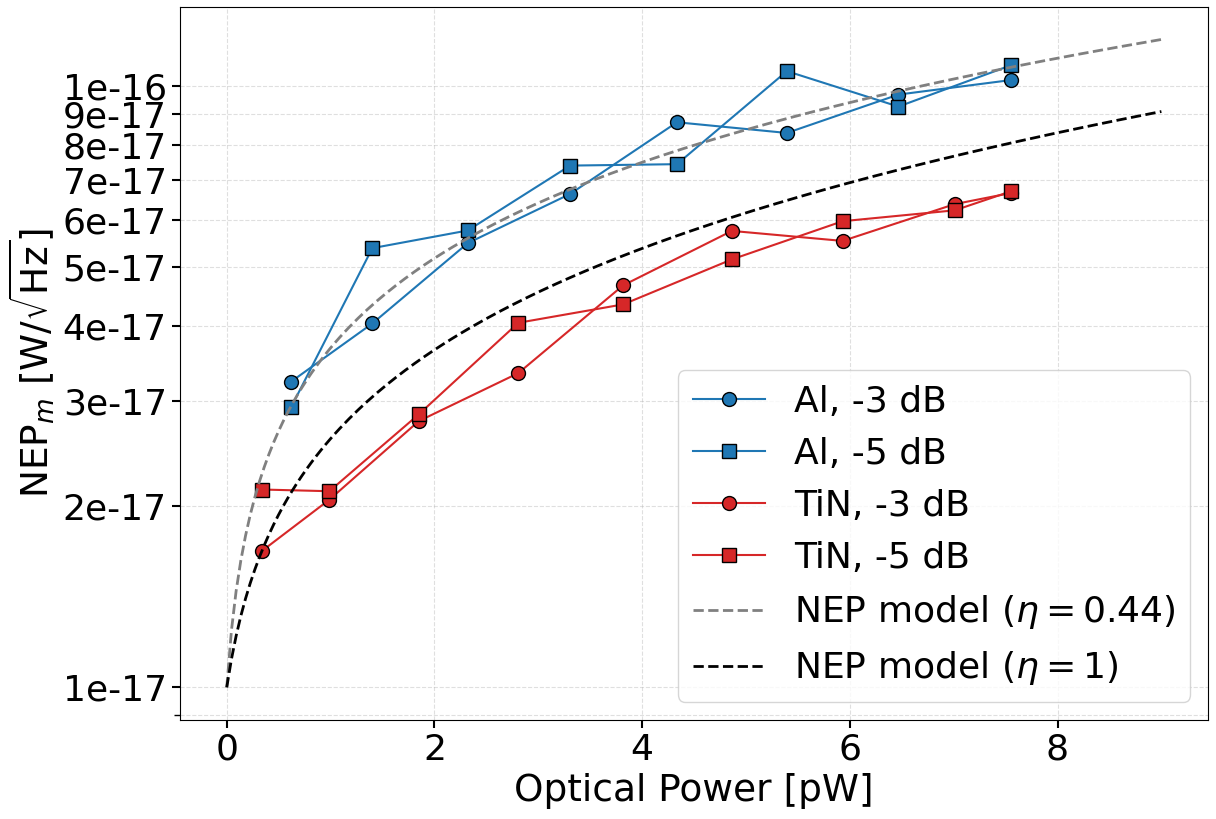}

    \caption{Top: Noise Equivalent Power (NEP) fitted with TLS contribution, modeled as a power-law spectrum, $\frac{1}{f^{-C}}$, and photon noise contribution, which appears flat at intermediate frequencies. Left: NEP for a TiN KID with the readout power at $3$\,dB from bifurcation. Right: NEP for an Al KID with $3$\,dB from bifurcation. Bottom: Detector NEP averaged at 100\,Hz over 20\,Hz and plotted versus estimated optical power for both TiN and Al 280\,GHz detectors. Two tone power levels are shown. For each optical loading, the optimal readout power is chosen to be 3\,dB below the bifurcation power. Additional offsets from this optimal power are included to illustrate the degradation in noise performance at suboptimal readout settings. We find that modest reductions in readout power do not significantly degrade the noise; however, at lower powers the amplitude noise increases and can dominate over the detector noise. Also shown are theoretical NEP shot noise models corresponding to 100\% and 44\% optical efficiencies.}
    \vspace{-1em}
    \label{fig:nep_vs_optical_power}
\end{figure}

\subsection{Optical Responsivity}  

To characterize optical responsivity, the cold load temperature was varied from 4\,K to 18\,K in 2\,K increments. At each temperature setpoint, data collection proceeded in three stages:  
(1) a broadband sweep to identify approximate resonant frequencies,  
(2) a narrowband target sweep around each resonance, and  
(3) timestream acquisition at the chosen tone frequencies and powers.  

All data (broadband sweep, target sweep, and timestream) are recorded in in-phase (I) and quadrature (Q) components, which are converted to amplitude and phase. For the resonant frequency $f_0$,  we use the frequency at which the transmission amplitude $|S_{21}(f)|$ reaches its minimum.

From the refined tone placements at each optical loading step, the fractional frequency shift $\Delta f / f_0$ is determined. The responsivity, $R$, is extracted from the slope of the $\Delta f / f_0$ versus optical power curve according to Equation~\ref{eq:responsivity}, taking into account the nonlinearity in responsivity where applicable. The optical responsivity $R$ is calculated as:  
\begin{equation}  
    R = \frac{\Delta f / f_0}{\Delta P_{\mathrm{opt}}} ,
\label{eq:responsivity}
\end{equation}  
where $P_{\mathrm{opt}}$ is the optical power incident on the detectors\cite{Zmuidzinas2012}.
Fig.~\ref{fig:s21_optical_power} shows example $|S_{21}|$ responses of 280\,GHz TiN and Al detector resonators at several levels of optical loading. The resulting fractional frequency shift, $\Delta f / f_0$, as a function of optical power is plotted in Fig.~\ref{fig:responsivity_vs_optical_power}. 

Fig.~\ref{fig:responsivity_vs_optical_power} illustrates that the TiN detectors exhibit a steeper, approximately linear responsivity compared to Al, which shows a nonlinear response \cite{Hubmayr2015,vaskuri2025_ccat_al_mkids}. Estimated optical loading is calculated from the cold load given its temperature and filter transmission, with an emissivity of 1. This follows previous measurements, with this cold load having an improved aspect ratio of the pyramidal geometry \cite{Choi2018}.

After the responsivity is calculated, the timestream data are analyzed. The phase timestream is mapped to frequency by fitting the measured phase–frequency relation near resonance with a cubic polynomial. Applying this fit converts the phase timestream $\phi(t)$ into a frequency timestream $f(t)$.
The fractional frequency shift is then computed as
\begin{equation}
\frac{\delta f(t)}{f_{0}} = \frac{f(t)}{\langle f(t) \rangle} - 1,
\end{equation}
which tracks frequency fluctuations relative to the mean resonance frequency. In practice $f(t)$ is computed over 20\,s timestream segments, which sets an effective low-frequency cutoff in the analysis. From this timestream, the fractional frequency noise spectrum is obtained as
\begin{equation}
S_{xx}(f) = \langle \left| \tilde{x}(f) \right|^2 \rangle,
\end{equation}
where $x(t) = \delta f(t)/f_{0}$ and $\tilde{x}(f)$ is the Fourier transform of $x(t)$.
In practice, $S_{xx}(f)$ is estimated using the Welch method (\texttt{scipy.signal.welch}),
which averages the power spectral density over overlapping segments of the timestream.

In addition, low-frequency noise components that are correlated across multiple detectors are identified and removed. This correlated noise, which is most prominent below 10\,Hz, may be attributed to common-mode fluctuations in the bath and cold load temperatures. The removal of this component reduces excess low-frequency noise specific to our testbed while preserving the uncorrelated detector noise used for their characterization. In the field, we expect the low-frequency noise to be dominated by the atmospheric fluctuations. No additional baseline corrections are applied in the frequency domain, and data below the effective cutoff are excluded from quantitative power spectral density (PSD) analysis.

The Noise Equivalent Power (NEP) is then calculated as:
\begin{equation}
\mathrm{NEP}(f) = \frac{\sqrt{S_{xx}(f)}}{R}.
\end{equation}
The responsivity used in this calculation is obtained by fitting the measured response as a function of optical power with a power-law model, as shown in Fig.~\ref{fig:responsivity_vs_optical_power}. The timestream is then rotated in the IQ plane to separate amplitude and phase directions, allowing the amplitude noise to be calculated. This amplitude contribution is subtracted at the level of the fractional frequency noise spectrum, $S_{xx}(f)$, to isolate the detector noise spectrum. The detector NEP is then computed from this corrected $S_{xx}(f)$. The NEP points are evaluated at the optimal readout power for each optical loading, defined as 3\,dB below the bifurcation power.

Fig.~\ref{fig:nep_vs_optical_power} (top) shows the total NEP spectral density, fitted using a noise model with a two level system (TLS) $\left(\frac{1}{f}\right)$ noise component and a white (photon) noise component. The extracted TLS slope and photon noise levels are consistent with typical observed values in superconducting detectors. A quasiparticle recombination roll-off is expected at higher frequencies due to the finite quasiparticle lifetime in the superconducting film. For the Al devices, quasiparticle lifetimes of up to $420\,\mu s$ are expected under similar operating conditions \cite{vaskuri2025_ccat_al_mkids}; however, the measured PSD does not extend to sufficiently high frequencies to constrain this feature and it is therefore not included in the fit. The low-frequency noise likely contains contributions from both detector-related processes and common-mode loading fluctuations. Fig.~\ref{fig:nep_vs_optical_power} (bottom) presents the NEP, averaged at 100\,Hz over a bandwidth of 20\,Hz, as a function of optical loading for both the 280\,GHz TiN and Al detectors biased at two tone powers. These results demonstrate how responsivity and noise together determine detector sensitivity and highlight the influence of both optical and readout power on the overall noise floor.

Using the standard NEP model, including photon noise and quasiparticle generation-recombination noise \cite{Zmuidzinas2003ThermalNoise, Hubmayr2015, vaskuri2025_ccat_al_mkids}, we infer optical efficiencies of $\sim$44\% for the Al detectors and $>$100\% for the TiN detectors. The Al efficiency is lower than the $\sim$60\% reported in previous studies \cite{vaskuri2025_ccat_al_mkids}, while the TiN value is unphysical. These values likely reflect limitations of the current test configuration, including non-ideal optical coupling with a feedhorn that lacks choke structures, which may reduce the coupled optical power for Al and allow additional unaccounted direct absorption of stray photons in TiN. Residual light leaks, unaccounted thermal effects, and modified TiN detector properties resulting from post-fabrication inductance adjustments are also expected to contribute to systematic uncertainty in the optical efficiency estimate. The scope of this paper is to provide a controlled test of the loading- and tuning-dependent detector response in the current setup; final optical efficiencies for deployment devices are better represented by dedicated optical characterization measurements.

\begin{figure}[htbp]
    \centering
    \vspace{-0.2cm}
    \includegraphics[width=0.4\textwidth]{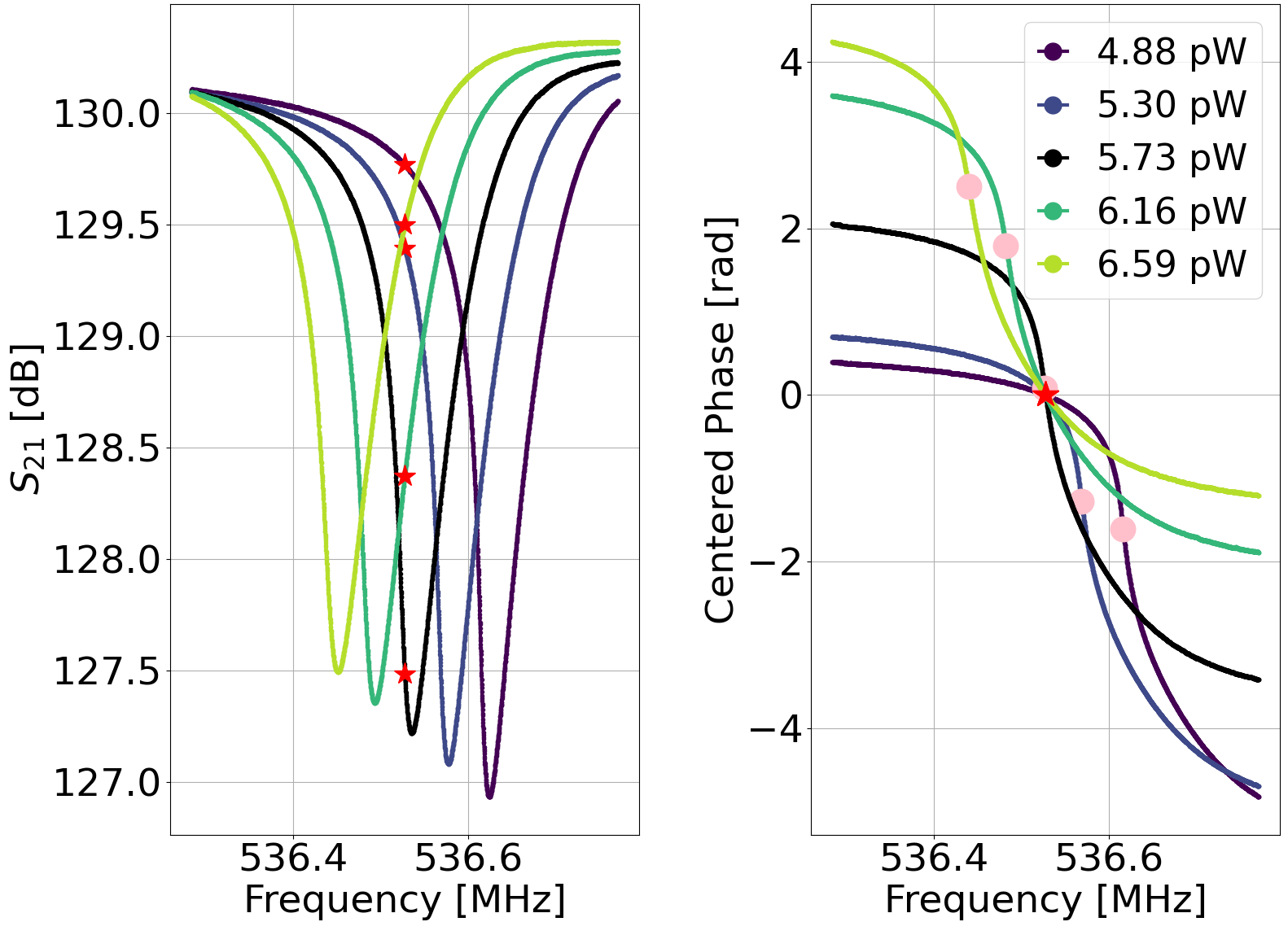}\\    
    \vspace{.2em}
    \includegraphics[width=0.44\textwidth]{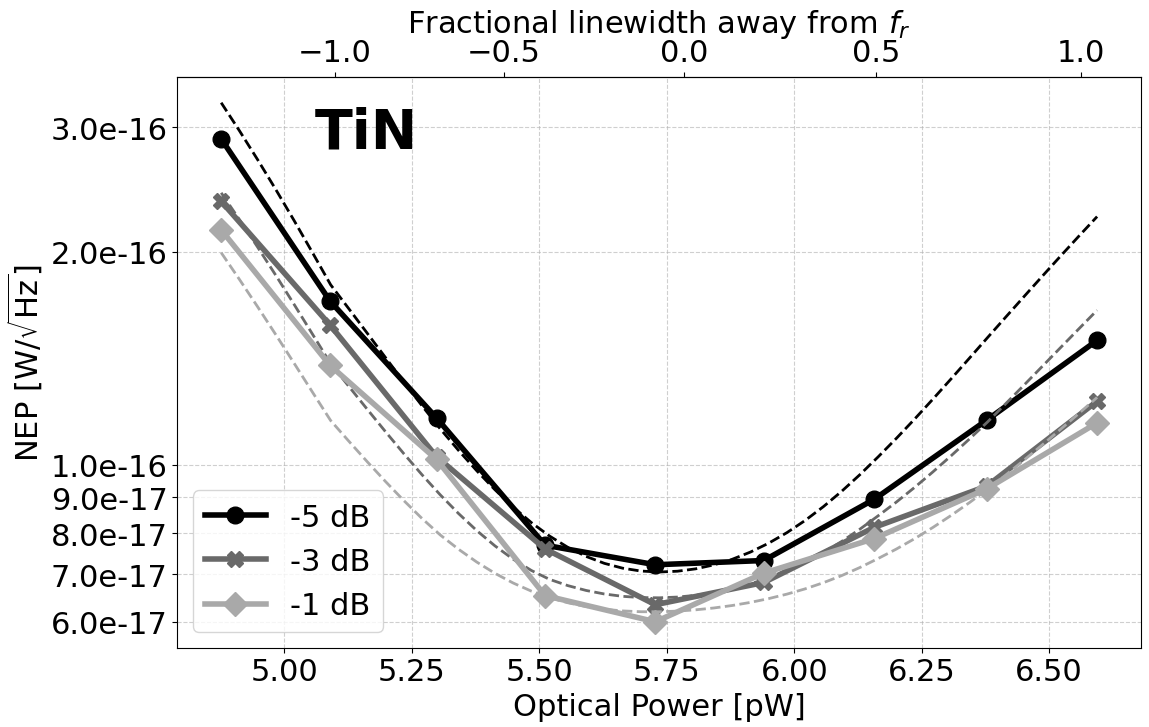}\\
    \vspace{.2em}
    \includegraphics[width=0.44\textwidth]{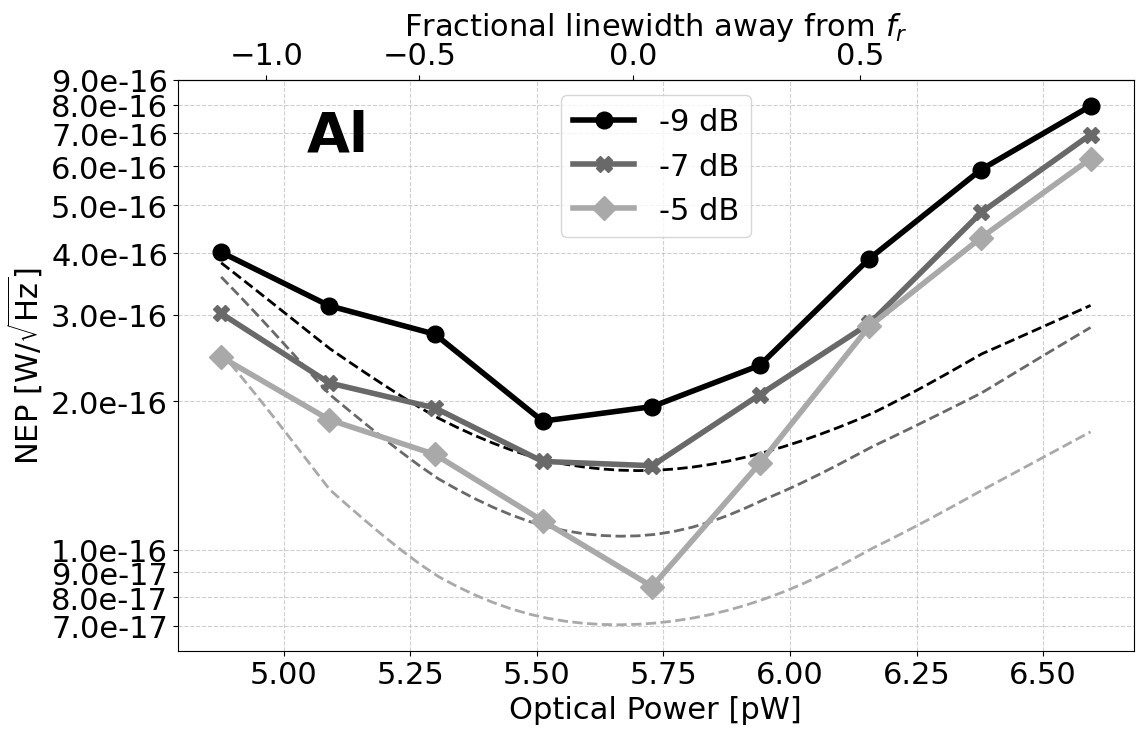}
    \vspace{-0.3cm}
    \caption{Top: $|S_{21}|$ and phase response of a representative TiN KID resonator, illustrating tone placement (red star) and the resulting detuning as the cold load temperature is varied from 14\,K to 17\,K, corresponding to the range of optical powers shown. The phase plot shows that the readout tone is initially placed on the black resonance at the point of maximum phase response. The pink markers indicate the shifted locations of maximum phase response as the resonance detunes with increasing optical loading. Middle (TiN KID) and Bottom (Al KID): NEP (averaged around 100\,Hz) as a function of estimated optical power during drift, for multiple tone power levels. Dashed lines show  models based on enhanced amplifier noise due to resonator detuning, as described in the text. The noise level increases by up to 30 -- 40\,\% with $\sim$0.2\,pW optical loading changes.}
    \vspace{-1.5em}
    \label{fig:drift_nep_vs_power}
\end{figure}
\subsection{Optical Loading Drift}  

To evaluate the effects of drifts in optical power expected during observations, we perform a separate test using the same data-taking procedure described above. The tone placement is fixed at the frequency corresponding to the maximum phase response, i.e., the point where $\mathrm{d}\phi/\mathrm{d}f$ is maximized. The $\mathrm{d}\phi/\mathrm{d}f$ value is calculated when the cold load is at 15.6\,K and kept fixed while changing the cold load temperature, mimicking a loading variation. The tone placement is indicated by the red stars in the $|S_{21}|$ and phase plots in Fig.~\ref{fig:drift_nep_vs_power} (top). In this configuration, the temperature of the cold load is shifted from 14\,K to 17\,K in 0.2\,K steps while the readout tone is fixed at the 15.6\,K resonance frequency value.  

Fig.~\ref{fig:drift_nep_vs_power} (top) shows an example of the response of a resonator $|S_{21}|$ and phase under these conditions, illustrating how detuning develops as the resonance changes with optical power. Amplitude and phase variations are recorded throughout the measurement and analyzed following the same procedure as in the responsivity measurement, including phase fitting, fractional frequency noise estimation, and NEP calculation.

When the readout tone is displaced from resonance by a fractional detuning $x = (f-f_r)/f_r$, the resonator transmission is described by the $S_{21}$ transfer function \cite{Swenson2013}:
\begin{equation}
    S_{21}(x) = 1 - \frac{Q_r}{Q_c} \cdot \frac{1}{1 + 2iQ_r x}
\end{equation}
where $Q_r$ and $Q_c$ are the total and coupling quality factors respectively, calculated by fitting the IQ circle. As optical loading shifts the resonance, the tone detunes from the point of maximum phase response, reducing the effective responsivity and increasing the contribution of amplifier-induced amplitude noise to the total NEP.

The amplifier-induced amplitude noise contribution is modeled using the resonator transfer function and phase response slope. Following \cite{Zmuidzinas2012, Swenson2013}, the amplifier noise spectral density is written as $\varepsilon = 4k_B T_{\mathrm{amp}} Z$ per unit bandwidth. Including the detuning dependence through $S_{21}(x)$ and the phase slope $\mathrm{d}\phi/\mathrm{d}x$, evaluated using fitted $Q_r$ and $Q_c$, gives the amplifier-noise contribution:
\begin{equation}
    \mathrm{NEP}_{\mathrm{amp}}(f) =
    \frac{1}{\mathcal{R}}
    \sqrt{\frac{k_B T_{\mathrm{amp}}}{P_{\mathrm{readout}}}}
    \cdot
    \frac{1}{|S_{21}(x)| \cdot |\mathrm{d}\phi/\mathrm{d}x|}
\end{equation}
where $\mathcal{R} = \mathrm{d}(\delta f/f_r)/\mathrm{d}P$ is the normalized responsivity and $P_{\mathrm{readout}}$ is the microwave readout power at the resonator.

The middle and bottom panels in Fig.~\ref{fig:drift_nep_vs_power} show the resulting total NEP measured (at 100\,Hz, averaged over a bandwidth of 20\,Hz) as a function of optical power for multiple tone readout powers during changes in cold load temperature. These results show that even moderate fluctuations in optical loading---on the order of 0.1--0.2\,pW---can cause up to a 40\,\% increase in noise, illustrating the sensitivity of KIDs to tone placement. The smooth lines represent the model from the quadrature sum of the amplifier, photon, and quasiparticle generation--recombination noise terms:
\begin{equation}
    \mathrm{NEP}_{\mathrm{tot}} =
    \sqrt{
    \mathrm{NEP}_{\mathrm{amp}}^2 +
    \mathrm{NEP}_{\mathrm{ph}}^2 +
    \mathrm{NEP}_{\mathrm{GR}}^2
    }
\end{equation}
The photon noise contribution, $\mathrm{NEP}_{\mathrm{ph}}$, is calculated following the formalism presented in \cite{Zmuidzinas2003ThermalNoise}, while $\mathrm{NEP}_{\mathrm{GR}}$ accounts for quasiparticle generation--recombination noise\cite{Zmuidzinas2012}.

Previous PWV fluctuations measured by APEX can range up to 0.1\,mm at 30\,min time scales, which corresponds to ~0.4\,pW of power variation\cite{Morris2025Atmospheric}. We expect a higher level of variations at the higher observing frequencies of CCAT. Using a simulated scan strategy corresponding to $\sim$\,$4^\circ$ elevation sweeps, we estimate atmospheric loading variations of approximately 0.1--0.2\,pW across a single scan, consistent with the fluctuation levels considered here \cite{paine_ammodel, Choi2020, ccat_sensitivity_calculator}. These estimates motivate further work in situ to determine the required cadence of detector retuning and minimize the noise penalty associated with detuning.

\section{Conclusion}

We have presented laboratory measurements of 280\,GHz TiN and Al KIDs fabricated alongside the first arrays to be deployed on the Prime-Cam receiver for the CCAT Observatory in 2026. These measurements provide a controlled testbed to evaluate detector performance and inform improvements to the full focal plane arrays. Using a cold load, we measured optical responsivity, noise performance, and NEP degradation due to optical loading drifts expected during observations. Despite the non-optimal laboratory configuration and use of sample pixels that differ from the final deployed designs, these measurements characterize the potential noise penalty associated with readout-tone detuning under varying optical loading, inform tuning and calibration strategy, and motivate the development of tone-tracking readout systems. 

The measured Al detector noise and responsivity are largely consistent with previous measurements of similar devices \cite{vaskuri2025_ccat_al_mkids}. While the Al devices are representative of the final detector design, the TiN pixels characterized here were fabricated prior to design updates made for the final arrays and exhibited higher-than-expected responsivity. Based on these measurements, the final TiN design incorporates a reduced inductor volume, which increases the quality factor and inductance and lowers the responsivity into the intended operating range.

We characterized the degradation of NEP with decreasing readout power, highlighting the importance of tone placement and readout optimization in large multiplexed arrays. At 0.5 linewidth detuning, corresponding to an optical power change of approximately 0.2\,pW, the NEP increased by a factor of 1.2--1.4. Detector response was also found to be non-linear, reinforcing the need for accurate responsivity models and lookup tables for calibration under varying loading conditions. This non-linearity is particularly pronounced in the Al arrays, requiring tighter control of tone placement and more extensive pre-deployment characterization, including responsivity as a function of optical loading and readout power.

Our results directly inform deployment and calibration strategies. Initial on-sky calibration will use celestial sources to measure relative detector gain as a function of elevation and atmospheric loading (PWV). The observed detector non-linearity requires these calibrations to be incorporated through detector-specific responsivity models, and we are developing analysis pipelines that directly include these characterization products to ensure stable performance when scaling to kilopixel arrays \cite{patel2025ccatreadout10000280}. Together, these measurements demonstrate the importance of thorough laboratory characterization prior to deployment and will guide detector design, calibration procedures, readout optimization, and retuning cadence under changing atmospheric conditions. The methods developed here will also be extended to future 350\,GHz and 850\,GHz Prime-Cam arrays to support the sensitivity and stability requirements for submillimeter cosmology and astrophysics.

\section*{Acknowledgments}
The CCAT project, FYST and Prime-Cam instrument have been supported by generous contributions from the Fred M. Young, Jr. Charitable Trust, Cornell University, and the Canada Foundation for Innovation and the Provinces of Ontario, Alberta, and British Columbia. The construction of EoR-Spec is supported by NSF grant AST-2009767. The construction of the 350 GHz instrument module for Prime-Cam is supported by NSF grant AST-2117631. The construction of the FYST telescope was supported by the Gro{\ss}ger{\"a}te-Programm of the German Science Foundation (Deutsche Forschungsgemeinschaft, DFG) under grant INST 216/733-1 FUGG, as well as funding from Universit{\"a}t zu K{\"o}ln, Universit{\"a}t Bonn and the Max Planck Institut f{\"u}r Astrophysik, Garching.

\bibliographystyle{IEEEtran}
\bibliography{refrences}

\vfill

\end{document}